\documentclass[conference]{IEEEtran}
\IEEEoverridecommandlockouts
\usepackage[utf8]{inputenc}
\usepackage{cite}
\usepackage{amsmath,amssymb,amsfonts}
\usepackage{algorithmic}
\usepackage{float}
\usepackage{color}
\usepackage{graphicx}
\usepackage{hyperref}
\usepackage{textcomp}
\usepackage{booktabs}
\usepackage{multirow}
\usepackage{colortbl}
\usepackage{tikz}
\usepackage{booktabs}

\title{Bit-Width-Aware Design Environment for Few-Shot Learning on Edge AI Hardware}

\author{%
    \IEEEauthorblockN{R. Kanda\IEEEauthorrefmark{1}\IEEEauthorrefmark{2}, H. L. Blevec\IEEEauthorrefmark{3}, N. Onizawa\IEEEauthorrefmark{1}, M. Leonardon\IEEEauthorrefmark{3}, V. Gripon\IEEEauthorrefmark{3} and T. Hanyu\IEEEauthorrefmark{1}}
    \IEEEauthorblockA{\IEEEauthorrefmark{1}Research Institute of Electrical Communication, Tohoku University, Japan}
    \IEEEauthorblockA{\IEEEauthorrefmark{2}Graduate School of Engineering, Tohoku University, Japan}
	\IEEEauthorblockA{\IEEEauthorrefmark{3}IMT Atlantique, Lab-STICC, UMR CNRS 6285, F-29238 Brest, France}
    Email: kanda.ryosuke.q3@dc.tohoku.ac.jp
}

\begin{document}
\maketitle

\begin{abstract}
    In this study, we propose an implementation methodology of real-time few-shot learning on tiny FPGA SoCs such as the PYNQ-Z1 board with arbitrary fixed-point bit-widths. 
    Tensil-based conventional design environments limited hardware implementations to fixed-point bit-widths of 16 or 32 bits. 
    To address this, we adopt the FINN framework, enabling implementations with arbitrary bit-widths. 
    Several customizations and minor adjustments are made, including:
    1.Optimization of Transpose nodes to resolve data format mismatches,
    2.Addition of handling for converting the final ``reduce mean'' operation to Global Average Pooling (GAP).
    These adjustments allow us to reduce the bit-width while maintaining the same accuracy as the conventional realization, and achieve approximately twice the throughput in evaluations using CIFAR-10 dataset. 

    \end{abstract}



\section{Introduction}
As Artificial Intelligence (AI) technology continues to evolve, the computational demands and energy consumption of traditional AI learning methods have escalated, creating an urgent need for innovative solutions. 
In response, research on edge AI hardware powered by Few-Shot Learning (FSL)\cite{wang2020generalizing}, which fundamentally diverges from conventional neural network learning, has gained increasing attention.
Few-shot learning has recently become a focal point in research areas such as natural language processing and image recognition. 
By enabling models to learn from limited datasets and generalize to new, unseen data, FSL offers the potential for more efficient, human-like learning, especially in situations where data availability is limited.

In this study, we refer to PEFSL (A Pipeline for Embedded Few-Shot Learning)\cite{PEFSL} as a representative example of few-shot learning implementation for edge AI. 
The PEFSL framework offers a complete pipeline covering pre-training, hardware synthesis, and the deployment of few-shot learning applications on FPGAs, using the PYNQ-Z1 board and the Tensil framework to carry out image recognition AI tasks.

However, PEFSL currently faces several challenges, one of which is the mismatch between pre-training, which is performed using floating-point operations on GPUs, and hardware implementation, which is conducted using fixed-point operations. 
Previous research\cite{mypaper} has addressed this issue by applying quantization modules to perform pre-training with fixed-point operations at arbitrary bit-widths.
On the other hand, Tensil imposes restrictions on the supported bit-widths, presenting challenges from the perspective of flexible hardware design. 
In this paper, we describe a design environment that leverages FINN\cite{finn}\cite{blott2018finn}, in contrast to Tensil, to enable hardware implementations with arbitrary bit-widths. 
FINN is an open-source framework for building, training, and deploying quantized neural networks on FPGAs, supporting a wide range of embedded applications.
This implementation allows for the unification of numerical representations between pre-training and hardware deployment, ensuring consistency in accuracy across the entire design flow and contributing to the efficiency of accuracy validation. 
Additionally, evaluations on the PYNQ-Z1 board, equipped with the Zynq Z-7020, demonstrate approximately double the throughput compared to the conventional design environment.

\begin{figure}[t]
	\centering
	\includegraphics[width=0.9\columnwidth]{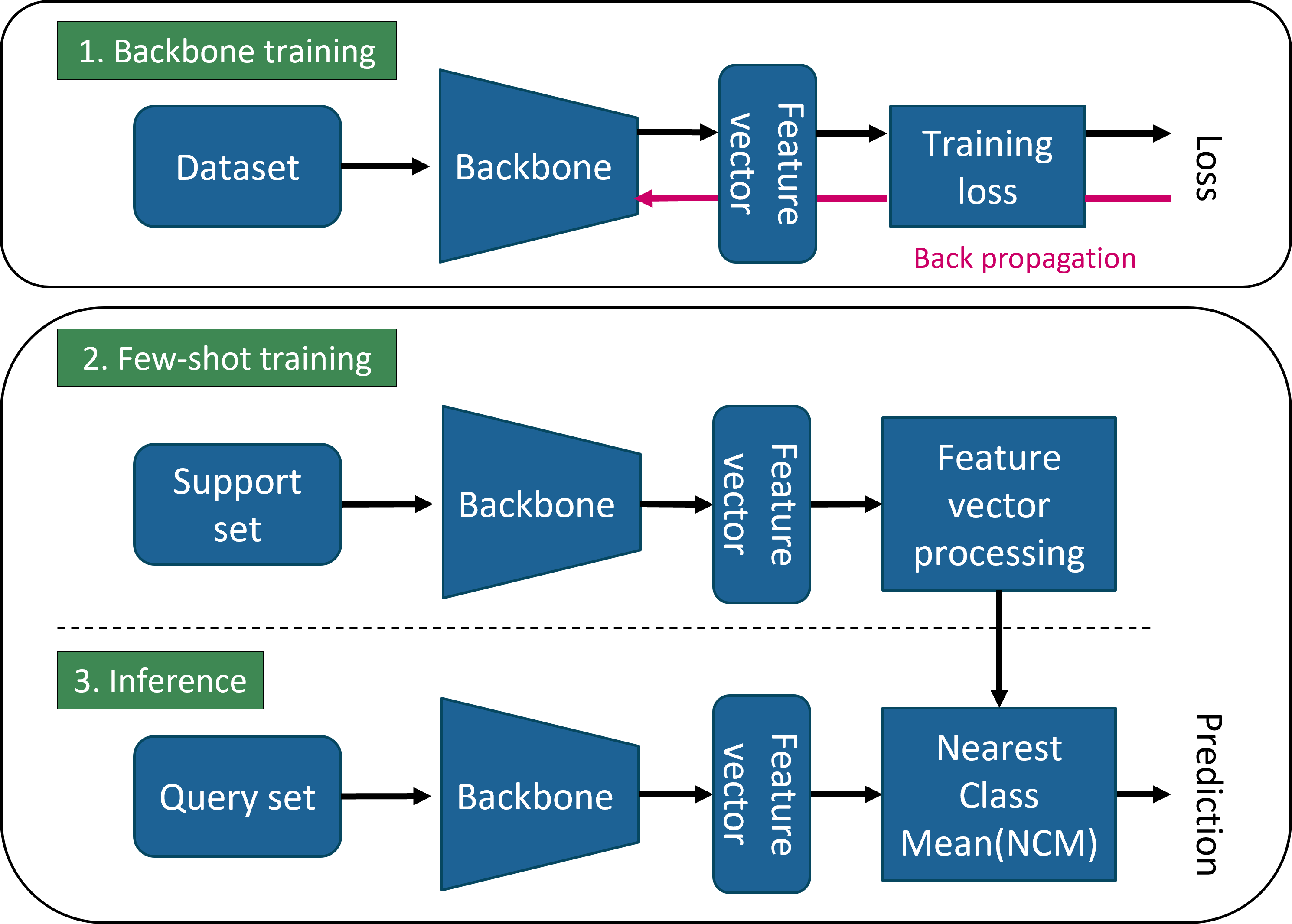}
	\caption{Few-shot learning consists of three main steps: 
    (1) training a backbone network to extract feature vectors from a dataset through backpropagation, 
    (2) using the pretrained backbone to generate feature vectors from a support set and then training a simple classifier based on them, and 
    (3) applying the trained backbone to a query set for inference, where classification is performed using a nearest class mean (NCM) approach.}
	\label{fsl_flow}
\end{figure}

\section{Framework for few-shot learning}
\subsection{Few-Shot Learning}
Few-shot learning is a learning approach aimed at learning from a small number of samples. 
In few-shot learning, two types of data samples are used: the support set and the query set. 
The support set is a small number of labeled samples from which the model is trained, and the query set consists of new samples used to evaluate the model's predictive ability. 
Few-shot learning incorporates the concepts of ``way'' and ``shot,'' where ``way'' refers to the number of classes to classify, and ``shot'' refers to the number of training samples per class. 
These concepts guide how the model learns from limited labeled data in the support set and evaluates its performance on new data.

The learning and inference flow of few-shot learning typically involves three stages: backbone training, learning from a few samples, and inference, as shown in Fig. \ref{fsl_flow}. 
The models and algorithms used in this study are based on the EASY method \cite{bendou2022easy}, which is known for its simplicity and high performance in few-shot learning tasks.

\subsubsection{Backbone training}
Backbones are trained on large datasets using deep learning models such as Convolutional Neural Networks (CNNs)\cite{he2015deep} to create generalized feature extractors.
However, it doesn't include the image classes of the support set to be trained in the next step.

\subsubsection{Learning from a few samples}
The pre-trained backbone is frozen. 
The learning data (support) for few-shot is passed through the backbone to extract features, and training is conducted using a simple classifier.

\subsubsection{Inference}
Finally, an image to be classified is first transformed into features using the backbone, then fed to the classifier for generating an output decision.

\subsection{PEFSL}
As stated in the introduction, this study is based on A Pipeline for Embedded Few-Shot Learning (PEFSL), an example of edge AI implementation. 
PEFSL provides a pipeline for executing few-shot learning on hardware using the Tensil framework, processing the weight data, which has been validated and evaluated through backbone learning and few-shot processing, for hardware implementation.

However, this pipeline faces several challenges, one of which is the limitation on the supported bit-widths for implementation in Tensil.
Specifically, only fixed-point bit-widths of 16 or 32 bits can be specified, which may result in excessive bit-widths relative to the required precision for certain use cases. 
Therefore, this study aims to develop a design environment that enables implementations with arbitrary bit-widths by utilizing a framework different from Tensil.

\begin{figure}[t]
	\centering
	\includegraphics[width=0.9\columnwidth]{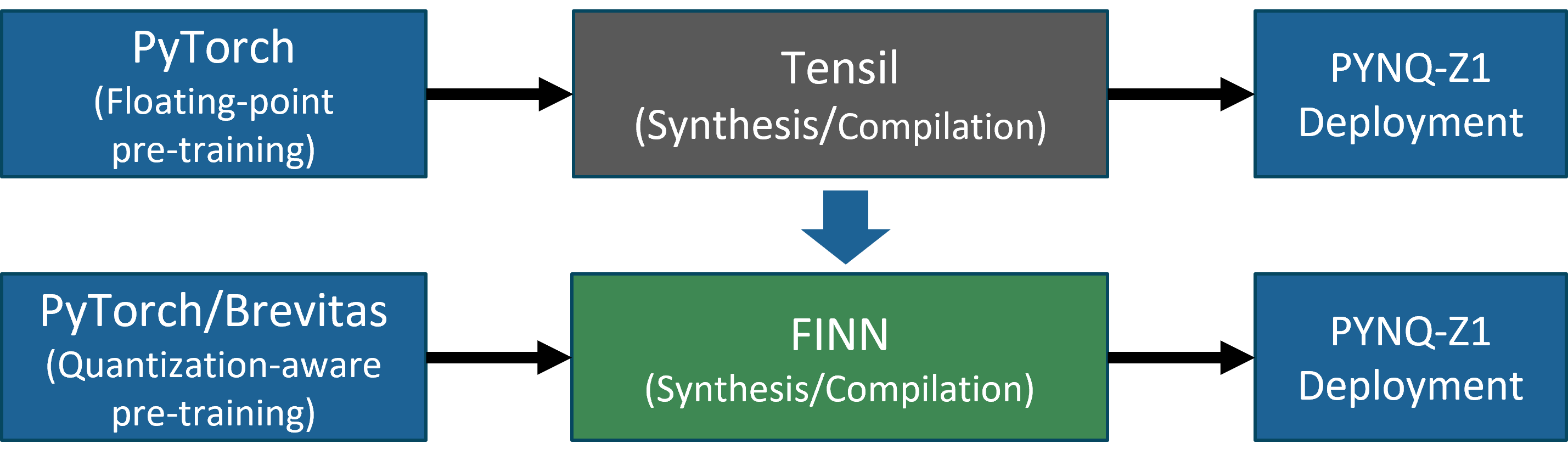}
	\caption{In this study, we replace the previous hardware conversion method using Tensil with FINN. 
             Starting from PyTorch-based pre-training (either floating-point or quantization-aware with Brevitas), models are synthesized through FINN for deployment on the PYNQ-Z1 board. 
             This approach leverages FINN’s efficient dataflow architecture, contrasting with the sequential processing style of Tensil.}
	\label{compare}
\end{figure}

\begin{figure}[t]
	\centering
	\includegraphics[width=0.9\columnwidth]{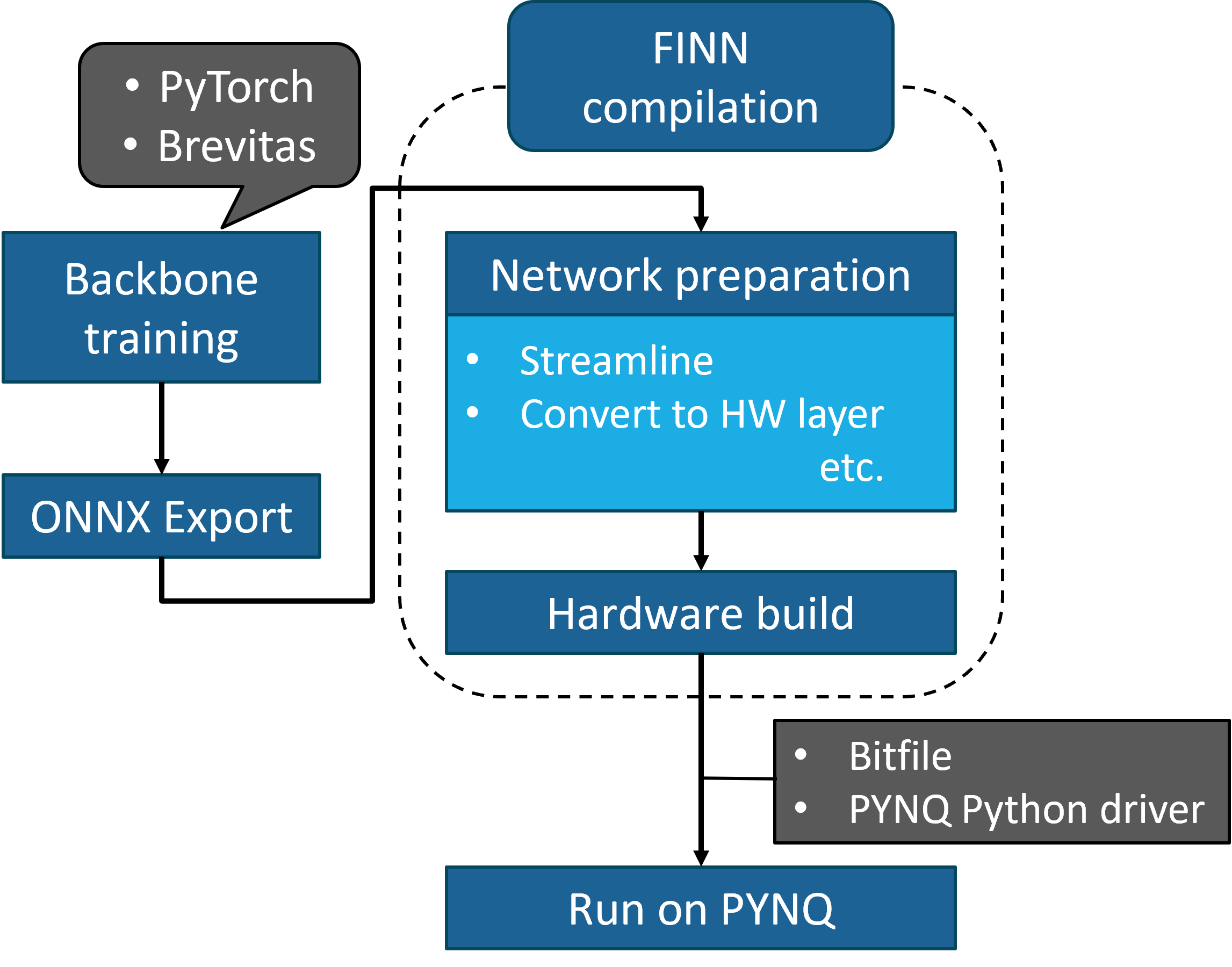}
	\caption{Flow from model training to hardware: A quantized model is generated using PyTorch with Brevitas \cite{brevitas}, trained to specific bit-widths suitable for FPGA deployment. 
             The trained model is exported as an ONNX file, which is then processed by FINN. 
             Through this process, FINN applies necessary transformations to produce the final bitfile and runtime driver for the FPGA.}
	\label{flow}
\end{figure}

\begin{table*}[t]
    \centering
    \caption{Architectural Comparison between FINN and Tensil}
    \label{tab:compare}
    \renewcommand{\arraystretch}{1.5} 
    \begin{tabular}{c||c|c}
        \hline
        \textbf{Category} & \textbf{Tensil} & \textbf{FINN} \\ \hline
        Structure & Matrix operations with a systolic array architecture & Streaming processing, layer-wise design with HLS/RTL \\ \hline
        Weights stored in & DRAM & BRAM \\ \hline
        Bit-width & Limited (fixed at 16 or 32 bits) & Flexible (allowing for arbitrary bit-widths) \\ \hline
        Latency & Can be higher due to DRAM access overhead & Low latency \\
        \hline
    \end{tabular}
\end{table*}

\section{FINN Customizations for Few-Shot Learning}
\subsection{Overview}
In this study, we adopted the FINN framework. 
FINN is an open-source framework provided by AMD, offering relaxed constraints on bit-widths and neural network architectures, enabling flexible hardware design. 
Additionally, since Brevitas, a quantization module used during pre-training to generate quantized models, is supported by FINN, the framework is well-suited for building a comprehensive design environment.

Fig.\ref{flow}. illustrates the flow from model training to hardware implementation. 
First, a quantized model with the specified bit-width is generated using PyTorch\cite{PyTorch} and Brevitas, and exported in ONNX\cite{onnx} format for use with FINN. 
FINN takes the ONNX model graph as its initial input and applies various transformations to map each node of the graph to the corresponding HLS/RTL layer. 
Through synthesis and other processes, the bitfile and runtime driver are automatically generated and implemented.

However, while FINN provides an excellent foundation for deploying neural networks on FPGAs, its standard build steps may not fully meet the specific requirements of the target application.
For example, the build steps provided in FINN’s tutorial for a simple Multi-layer Perceptron (MLP) cannot be directly applied to other architectures. 
To address this constraints, we customized the build steps for ResNet-9, the backbone model used for few-shot learning in this study. 
In particular, the ``Streamline'' and ``Convert to HW Layer'' steps in the Network Preparation phase, as shown in Fig. 3, are architecture-dependent, and we adjusted these steps accordingly. 
While FINN provides various transformation classes for converting model graphs for hardware deployment, which are customized to suit the specific architecture being used, we mainly utilized elements that were already available within the FINN framework, even though they were not part of the default tutorial flow.

\subsection{Tensil and FINN} 
Table \ref{tab:compare} describes the key differences between the implementations using Tensil and FINN. 
First, we outline the architectural distinctions between the two frameworks.
FINN leverages HLS/RTL to design each layer individually, connecting them through FIFO buffers to enable low-latency streaming processing. 
Additionally, FINN stores weights in BRAM (Block RAM), eliminating the overhead of DRAM access. 
In contrast, Tensil employs a systolic array architecture, efficiently parallelizing matrix operations to support large-scale models. 
However, Tensil stores both weights and activations in DRAM, potentially introducing a bottleneck due to memory access latency.

\subsection{Transpose Node Optimization}
Regarding the data format of input and output tensors, PyTorch commonly uses NCHW (Number of batches, channels, height, width), while FINN’s HLS/RTL libraries primarily employ NHWC. 
As a result, Transpose nodes are automatically inserted during the build process. 
However, to ensure the implementation functions as intended, these nodes must be handled appropriately.

\begin{figure}[t]
	\centering
	\includegraphics[width=0.9\columnwidth]{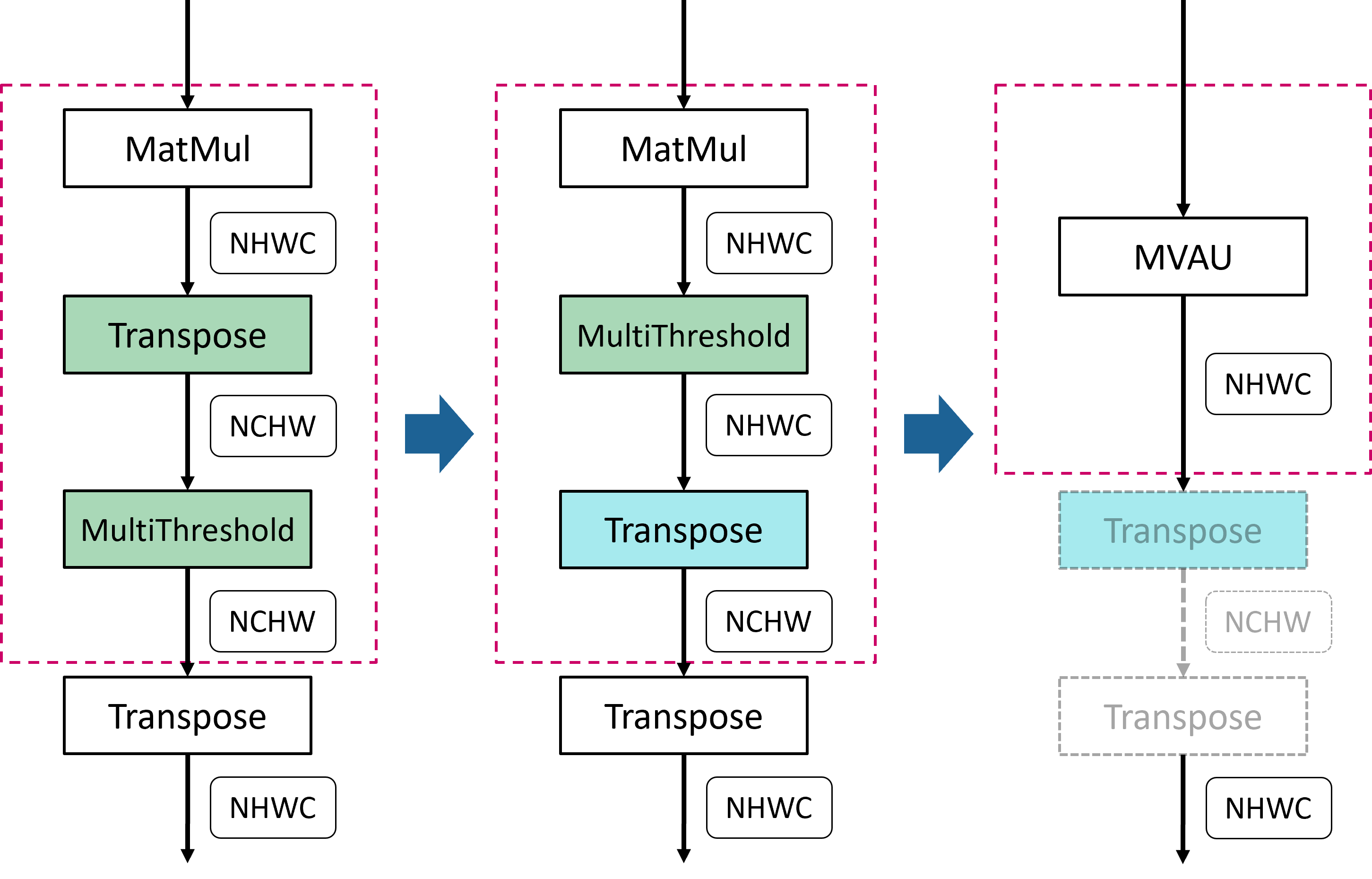}
	\caption{MatMul outputs in NHWC format, while MultiThreshold expects NCHW, requiring a Transpose operation in between. 
             This mismatch led to improper weight transfer to the MVAU, causing processing errors. 
             The issue was resolved by merging the nodes using AbsorbTransposeIntoMultiThreshold and inserting a Transpose afterward to ensure correct data flow for subsequent layers.
    }
	\label{noact}
\end{figure}

Fig.\ref{noact}. illustrates an overview of this process. Specifically, the Conv processing MatMul node outputs in NHWC format, while the MultiThreshold node expects input in NCHW format, necessitating the insertion of a Transpose node before MultiThreshold. 
This mismatch prevented the proper transfer of weights to the MVAU (Matrix Vector Activation Unit). 
To resolve this issue, we introduced the AbsorbTransposeIntoMultiThreshold transformation, merging the two nodes and inserting a Transpose node afterward.

\subsection{Reduce Mean and GAP Handling}
The model used in this study incorporates a ``reduce mean'' operation in its final layer, which performs spatial dimensionality reduction by applying the operation along the height and width axes. 
In previous design environments utilizing Tensil, the reduce mean operation could not be directly handled, requiring conversion to Global Average Pooling (GAP), which achieves an equivalent operation. 
Similarly, in FINN, processing through GAP is required, and we have added a custom transformation class specifically for GAP to accommodate this need, ensuring seamless conversion and accurate handling of the model's final layer.
The GlobalAccPool, a custom node in FINN, computes the cumulative sum along the spatial dimensions (height and width) of the feature maps, similar to GAP, but incorporates several optimizations to maximize resource efficiency. 
Instead of performing division within the class itself, it outputs the cumulative sum as is. 
The averaging is then achieved by applying scalar multiplication through a Mul node, effectively avoiding the computationally intensive division operation and ensuring more efficient processing.

In addition, we have tailored the build process for the ResNet-9 model by introducing transformation classes not included in the default build and rearranging the order of transformations as needed.

\section{Evaluation}
\subsection{Accuracy}
It has been confirmed from the findings of our previous work that the bit-widths traditionally used in hardware implementations are unnecessarily large for the required accuracy. 
This study evaluates bit-width configurations that can maintain the same level of accuracy as conventional implementations while reducing the bit-width size.
The model used in this study is ResNet-9, trained on the MiniImageNet dataset (resized to a resolution of 32×32). 
Accuracy evaluation was conducted using CIFAR-10\cite{cifar-10} dataset with a 5-way, 5-shot configuration. 
The results are summarized in Table \ref{tab:accuracy}.

The conventional 16-bit implementation achieved an accuracy of 62.78\%. 
Based on this result, we explored the possibility of reducing the bit-width while maintaining as much accuracy as possible. 
Taking into account the hardware resources of the PYNQ-Z1 board, we selected appropriate bit-widths. 
According to \cite{ducasse2021benchmarkingquantizedneuralnetworks}, activation layers tend to consume more resources, while reducing their bit-width has minimal impact on accuracy compared to convolutional layers. 
Based on these insights, we tested convolutional layers with 6-bit quantization (1 bit for the integer part and 5 bits for the fractional part) and activation layers with 4-bit quantization (2 bits for the integer part and 2 bits for the fractional part). 
The resulting accuracy was 59.70\%, demonstrating that the bit-width could be reduced while largely maintaining the original accuracy.

\begin{table}[t]
    \centering
    \caption{Accuracy on CIFAR-10 (5-shot)}
    \label{tab:accuracy}
    \begin{tabular}{cccccc}
    \toprule
    \multirow{2}*{Max bit-width} & \multicolumn{2}{c}{Conv layer} & \multicolumn{2}{c}{ReLU layer} & \multirow{2}*{Acc.} \\
     & int. & frac. & int. & frac. &  \\
    \midrule
    5 & 2 & 3 & 2 & 2 & 44.89 \\

    \hline
    \multicolumn{1}{|c}{6} & 1 & 5 & 2 & 2 & \multicolumn{1}{c|}{\bf{59.70}} \\
    \hline 

    6 & 3 & 3 & 3 & 3 & 44.72 \\
    8 & 4 & 4 & 4 & 4 & 60.92 \\
    10 & 5 & 5 & 5 & 5 & 62.58 \\
    12 & 6 & 6 & 6 & 6 & 62.69 \\
    14 & 7 & 7 & 7 & 7 & 62.47 \\
    \midrule
    16 & 8 & 8 & 8 & 8 & \textbf{62.78} \\
    \bottomrule
    \end{tabular}
\end{table}

\subsection{Implementation results}
Based on the results in Table \ref{tab:accuracy}, the model was built using a quantization of 6 bits (1 bit for the integer part and 5 bits for the fractional part) for the convolutional layers, and 4 bits (2 bits for the integer part and 2 bits for the fractional part) for the activation layers. 
The FPGA frequency has been set to 125 MHz.
The results are summarized in Table \ref{tab:resource}.

\begin{figure}[t]
	\centering
	\includegraphics[width=0.8\columnwidth]{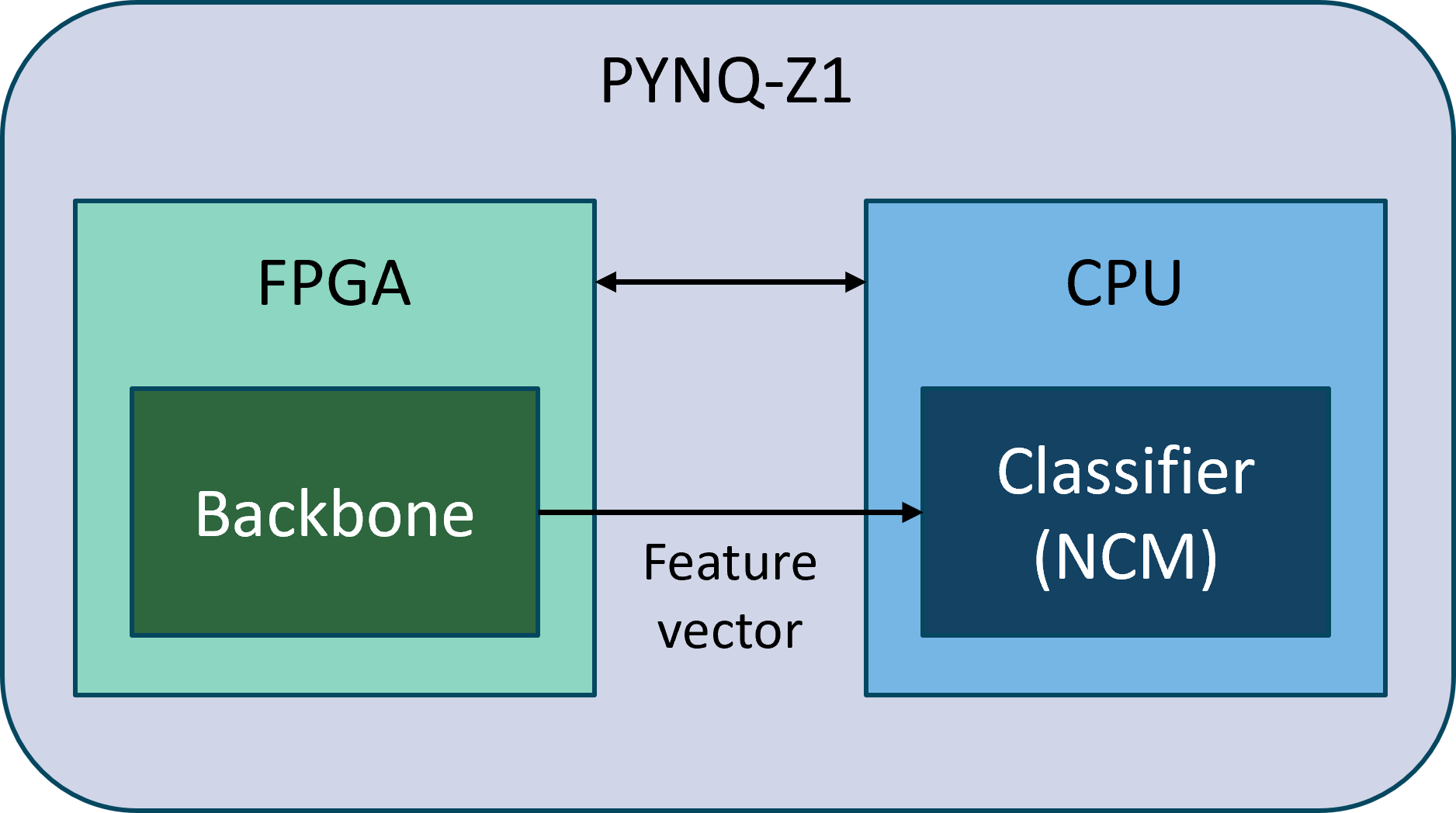}
	\caption{Overview of the implementation on the PYNQ-Z1 board. 
    The backbone network is deployed on the FPGA, which extracts feature vectors from input data. 
    These feature vectors are transferred to the CPU, where the NCM (Nearest Class Mean) classifier performs the final classification. }
	\label{overview}
\end{figure}

First, the DSP utilization is lower in FINN compared to PEFSL with Tensil. 
This is because Tensil’s systolic array heavily relies on DSP blocks to efficiently parallelize matrix operations. 
On the other hand, FINN increases the usage of LUTs and flip-flops (FFs) since each operation is directly implemented as a logic circuit. 
Additionally, BRAM usage is also higher in FINN, as weights are stored within the FPGA’s BRAM.
These architectural differences are reflected in the latency measurements. 
The implementation with FINN achieves a backbone inference latency of 16.3 ms on the PYNQ-Z1 FPGA (as shown in Fig.\ref{overview}.), which is approximately twice as fast as the conventional implementation. 
Since FINN stores the weights in BRAM, allowing direct access within the FPGA, it avoids the overhead associated with data transfers to and from DRAM, as seen in Tensil. 
As a result, the FINN-based implementation achieves shorter latency.
Furthermore, the implementation achieved a throughput of 61.5 fps, demonstrating high throughput performance. 
This enables real-time inference, confirming its effectiveness for edge AI applications.
%


\begin{table}[t]
	\centering
	\caption{CIFAR-10 inference on PYNQ-Z1 FPGA}
	\label{tab:resource}	
	\begin{tabular}{ccccccc}
	\toprule
    Work & Prec. & LUT & BRAM & FF & DSP & Latency \\
    & [bits] &  & [36kb] & & & [ms] \\
	\midrule
	PEFSL\cite{PEFSL} & 16 & 15667 & 59 & 9819 & 159 & 35.9  \\
	\bf{Ours} & 6 & 37263 & 131.5 & 44617 & 22 & 16.3  \\

	\bottomrule
	\end{tabular}
\end{table}

\section{Conclusion}
In this study, we proposed a design environment for the edge AI implementation of few-shot learning on the PYNQ-Z1 board using FINN. 
The proposed implementation maintains the same level of accuracy as conventional methods while achieving a latency of 16.3 ms. 
This achievement contributes to the development of high-performance real-time few-shot learning implementations.

Future work will focus on extending the FPGA implementation beyond the backbone, which is currently the only part implemented on the FPGA, by also offloading the classifier and other components currently handled by the CPU. 
This will further enhance performance and improve energy efficiency.

\section*{Acknowledgment}
This work was supported in part by JST CREST Grant Number JPMJCR19K.

\bibliographystyle{IEEEtran}
\bibliography{ISCAS}

@misc{he2015deep,
  title        = {Deep Residual Learning for Image Recognition},
  author       = {Kaiming He and Xiangyu Zhang and Shaoqing Ren and Jian Sun},
  note         = {arXiv:1512.03385},
  year         = {2015},
  eprint       = {1512.03385},
  archivePrefix= {arXiv},
  primaryClass = {cs.CV}
}

@misc{wang2020generalizing,
  title        = {Generalizing from a Few Examples: A Survey on Few-Shot Learning},
  author       = {Yaqing Wang and Quanming Yao and James Kwok and Lionel M. Ni},
  note         = {arXiv:1904.05046},
  year         = {2020},
  eprint       = {1904.05046},
  archivePrefix= {arXiv},
  primaryClass = {cs.LG}
}

@misc{bendou2022easy,
  title        = {{EASY}: Ensemble Augmented-Shot Y-shaped Learning: State-Of-The-Art Few-Shot Classification with Simple Ingredients},
  author       = {Yassir Bendou and Yuqing Hu and Raphael Lafargue and Giulia Lioi and Bastien Pasdeloup and Stéphane Pateux and Vincent Gripon},
  note         = {arXiv:2201.09699},
  year         = {2022},
  eprint       = {2201.09699},
  archivePrefix= {arXiv},
  primaryClass = {cs.LG}
}

@INPROCEEDINGS{PEFSL,
  author={Grativol, Lucas and Gauthier, Lubin and Léonardon, Mathieu and Morlier, Jérémy and Lavrard-Meyer, Antoine and Muller, Guillaume and Fresse, Virginie and Arzel, Matthieu},
  booktitle={2024 IEEE International Symposium on Circuits and Systems (ISCAS)}, 
  title={{PEFSL}: A deployment Pipeline for Embedded Few-Shot Learning on a FPGA SoC}, 
  year={2024},
  volume={},
  number={},
  pages={1-5},
  doi={10.1109/ISCAS58744.2024.10557995}}

@software{brevitas,
  author       = {Alessandro Pappalardo},
  title        = {Xilinx/brevitas},
  year         = {2023},
  publisher    = {Zenodo},
  doi          = {10.5281/zenodo.3333552},
  url          = {https://doi.org/10.5281/zenodo.3333552}
}

@inproceedings{finn,
  author = {Umuroglu, Yaman and Fraser, Nicholas J. and Gambardella, Giulio and 
            Blott, Michaela and Leong, Philip and Jahre, Magnus and Vissers, Kees},
  title = {{FINN}: A Framework for Fast, Scalable Binarized Neural Network Inference},
  booktitle = {Proceedings of the 2017 ACM/SIGDA International Symposium on Field-Programmable Gate Arrays},
  series = {FPGA '17},
  year = {2017},
  pages = {65--74},
  publisher = {ACM}
}

@article{blott2018finn,
  title = {{FINN-R}: An End-to-End Deep-Learning Framework for Fast Exploration of Quantized Neural Networks},
  author = {Blott, Michaela and Preu{\ss}er, Thomas B and Fraser, Nicholas J and Gambardella, Giulio and 
            O’brien, Kenneth and Umuroglu, Yaman and Leeser, Miriam and Vissers, Kees},
  journal = {ACM Transactions on Reconfigurable Technology and Systems (TRETS)},
  volume = {11},
  number = {3},
  pages = {1--23},
  year = {2018},
  publisher = {ACM New York, NY, USA}
}

@INPROCEEDINGS{mypaper,
  author={Kanda, R. and Onizawa, N. and Leonardon, M. and Gripon, V. and Hanyu, T.},
  booktitle={2024 IEEE 67th International Midwest Symposium on Circuits and Systems (MWSCAS)}, 
  title={Design Environment of Quantization-Aware Edge AI Hardware for Few-Shot Learning}, 
  year={2024},
  volume={},
  number={},
  pages={928-931},
  doi={10.1109/MWSCAS60917.2024.10658789}}

@misc{ducasse2021benchmarkingquantizedneuralnetworks,
      title={Benchmarking Quantized Neural Networks on FPGAs with FINN}, 
      author={Quentin Ducasse and Pascal Cotret and Loïc Lagadec and Robert Stewart},
      year={2021},
      eprint={2102.01341},
      archivePrefix={arXiv},
      primaryClass={cs.LG},
      url={https://arxiv.org/abs/2102.01341}, 
}

@misc{PyTorch,
      title={{PyTorch}: An Imperative Style, High-Performance Deep Learning Library}, 
      author={Adam Paszke and Sam Gross and Francisco Massa and Adam Lerer and James Bradbury and Gregory Chanan and Trevor Killeen and Zeming Lin and Natalia Gimelshein and Luca Antiga and Alban Desmaison and Andreas Köpf and Edward Yang and Zach DeVito and Martin Raison and Alykhan Tejani and Sasank Chilamkurthy and Benoit Steiner and Lu Fang and Junjie Bai and Soumith Chintala},
      year={2019},
      eprint={1912.01703},
      archivePrefix={arXiv},
      primaryClass={cs.LG},
      url={https://arxiv.org/abs/1912.01703}, 
}

@misc{onnx,
    author = {Bai, Junjie and Lu, Fang and Zhang, Ke and others},
    title = {ONNX: Open Neural Network Exchange},
    year = {2019},
    publisher = {GitHub},
    journal = {GitHub repository},
    howpublished = {\url{https://github.com/onnx/onnx}},
    commit = {94d238d96e3fb3a7ba34f03c284b9ad3516163be}
}

@inproceedings{cifar-10,
  title={Learning Multiple Layers of Features from Tiny Images},
  author={Alex Krizhevsky},
  year={2009},
  url={https://api.semanticscholar.org/CorpusID:18268744}
}

\end{document}